\documentclass[twocolumn]{aastex61}
\usepackage{amsmath,amstext}
\usepackage[T1]{fontenc}
\usepackage{apjfonts} 
\usepackage{graphicx}
\usepackage{subfigure}
\usepackage{xcolor}
\numberwithin{equation}{section}

\shorttitle{Dense Circumstellar Medium \& type II-P/IIL Supernova Light Curves}
\shortauthors{S. Das \& A. Ray}

\begin{document}




\title{Modelling type IIP/IIL supernovae interacting with recent episodic mass ejections from their presupernova stars with MESA \& SNEC}

\author{Sanskriti Das}\thanks{dassanskriti@gmail.com}
\affiliation{Department of Physics, Indian Institute of Technology Bombay, Mumbai 400076, India}
\affiliation{Department of Astronomy, The Ohio State University, 140 W 18$^{th}$ Avenue, Columbus, OH 43201}\thanks{Present address}
\author{Alak Ray}
\affiliation{Tata Institute of Fundamental Research, Colaba, Mumbai 400005, India}
\affiliation{Homi Bhabha Centre for Science Education (TIFR), Mankhurd, Mumbai 400088, India}\thanks{Present address}

\begin{abstract}
We show how dense compact discrete shells of circumstellar gas immediately outside the red supergiants affect the optical light curves of type II-P/II-L SNe taking the example of SN 2013ej. The earlier efforts in the literature had used an artificial circumstellar medium (CSM) stitched to the surface of an evolved star which had not gone through a phase of late-stage heavy mass loss, 
which in essence, is the source of the CSM to begin with. In contrast we allow enhanced mass loss rate from the modeled star during the $^{16}$O and \textcolor{black}{$^{28}$Si} burning stages and construct the CSM from the resulting mass-loss history in a self-consistent way. Once such evolved pre-SN stars are exploded, we find that the models with early interaction between the shock and the dense CSM reproduce the light curves far better than those without that mass loss and hence having no dense, nearby CSM. The required explosion energy for the progenitors with a dense CSM is reduced by almost a factor of two compared to those without the CSM. Our model, with a more realistic CSM profile and presupernova and explosion parameters, fits observed data much better throughout the rise, plateau and radioactive tail phases compared to previous studies. This points to an intermediate class of supernovae between type II-P/II-L and type II-n SNe with the characteristics of simultaneous UV and optical peak, slow decline after peak and a longer plateau. 
\end{abstract}

\section{Introduction}
Massive stars (with Zero Age Main Sequence (ZAMS) mass $>$ 8M$_\odot$) end their lives as core-collapse supernovae (CCSNe). Type-II SNe, a class of CCSNe are identified by the P-Cygni profile of hydrogen in their early spectra. Type II-P SNe comprise a large fraction ($\sim 48\%$) of nearby core-collapse SNe population within 60Mpc \citep{SmithN_etal2011}. These have pronounced plateaus in their visible band light curves that remain within $\sim$ 1 mag of maximum brightness for an extended period e.g. 60-100 rest frame days and is followed by exponential tail at late times \citep{Faran_etal2014a}. 
\textcolor{black}{Based on photometric and spectroscopic analysis, type II-L SNe are separately classified from II-P by their larger rise time, brighter peak, faster and linearly falling luminosity after peak and around $\sim$50 days, higher $H_\alpha$ velocity, larger ratio of $H_\alpha$ emission to absorption, slowly evolving $H_\beta$ velocity and bluer color curves with average decline rate $\beta^B >$ 3.5mag/100 days \citep{Barbon_etal1979,Patat_etal1993,Patat_etal1994,Arcavi_etal2012,Gutierrez_etal2014,Faran_etal2014b,Gall_etal2015}. At the same time, several studies with large samples show that types II-P and II-L SNe form a continuous and statistically indistinguishable class of CCSNe \citep{Young_etal1989,Anderson_etal2014,Sanders_etal2015}. 
Despite their quantitative differences in light curve patterns and spectroscopic features, different groups of researchers \citep{Grassberg_etal1971,Swartz_etal1991,Blinnikov_etal1993} have tried to find a range of progenitors that would lead to a continuous transition of the properties of type II-P to II-L explosions, in view of the possible variation of the plateau length and the associated parameters with the mass and radius of hydrogen envelope \citep{Litvinova_etal1983,Nomoto_etal1995}.
However, so far it has not been possible to simulate type II-L explosions as an extreme case of type II-P SNe \citep{MorozovaV_etal2015}. 
On the other hand Type II-n SNe \citep{Schlegel_1990} show in their early spectra narrow Hydrogen Balmer emission lines or P-cygni profiles on top of broad emission lines. A particularly interesting example SN1994W shows the presence of three components in their first three months consisting of narrow P-cygni lines with absorption minimum at 700 km/s, broad emission line with blue edge at $\sim$4000 km/s and broad smooth wings extended to at least 5000 km/s in H-$\alpha$ and H-$\beta$. \cite{Chugai_etal2004} attribute these components to a dense circumstellar envelope, a shocked, dense but cool gas confined right on top of the photosphere and the effects of Thompson scattering in the circumstellar envelope. The plateau-like light curve of the supernova modeled by hydrodynamic simulations showed that the pre-explosion kinematics of the circumstellar envelope for a high mass ($\sim$0.4 M$_\odot$) and kinetic energy ($\sim 2\times 10^{48}$ erg) of the envelope which must be ejected $\sim$ 1.5 years prior to the explosion. A close cousin of SN1994W is SN2011ht which also shows a plateau and subsequent exceptionally faint and steeply declining light curve in its nebular phase prompting \cite{Mauerhan_etal2013} to propose a new subclass of type II-nP interacting SNe with a plateau light curve phase. Typically type II-n SNe have large rise time ($\sim$20-50 days), very bright peaks of M$_V$ = -18.4$\pm$1.0 mag and a large range of decline rate spanning the range from flat plateau events like II-P SNe to rapidly decaying events like II-b SNe\citep{Kiewe_etal2012}. These works support the association of II-n SNe with LBV (Luminous Blue Variable) stars.   
}
\\
Type II-P/II-L and II-n SNe have been traditionally differentiated in separate categories due to their observed characteristics and inferred progenitor properties. Recently there has been consideration of the effect of dense circumstellar material immediately outside the progenitors on the early light curves of type II-P/II-L SNe \citep{Valenti_etal2015, MorozovaV_etal2017, Yaron_etal2017} and a subclass of \textit{moderately interacting SNe} has been identified. This subclass may comprise an intermediate class of SNe between type II-P/II-L and II-n SNe \citep{SmithN_etal2015} and their progenitors may fill the gap between observed ZAMS mass range of their respective progenitors \citep{Moriya_etal2011}. \\ 
Here we have evolved the stars in MESA: Modules for Experiments in Stellar Astrophysics\footnote{\url{http://mesa.sourceforge.net/index.html}} \citep{Pax2011,Pax2013,Pax2015} since their pre-ZAMS stage till the Fe core collapse with a history of enhanced mass loss rate in last few years, and constructed the CSM from the information of that episodic mass ejection during late stage evolution of the star. Our method is self-consistent in two ways. Firstly, since the simulation of enhanced mass loss over a timescale of few years reveals the impact on surface properties (mainly luminosity and radius) naturally from the stellar evolution (Fig.\ref{fig1}), the pre-SN progenitor carries the trace of the phenomenon with it. Secondly, the modeling of the CSM also accounts for these changes in the star, and thus the CSM profile becomes more realistic. This is in contrast with the earlier works on the effect of CSM on type II-P SNe where artificial CSM is designed assuming a constant ratio of mass loss rate and wind speed and the profile is stitched to the pre-SN star which is independently modeled by stellar evolution codes \textcolor{black}{(e.g. KEPLER or MESA)} without any huge mass loss history \citep{MorozovaV_etal2017, Moriya_etal2017}. We explode the progenitor with CSM using SNEC: SuperNova Explosion Code \footnote{\url{https://stellarcollapse.org/SNEC}}\citep{MorozovaV_etal15} and compare its light curve with the light curve of the progenitor with no excess mass loss history and hence no dense CSM in the immediate vicinity of the exploding star. We make a comparative study of the models best fitting the optical and NIR light curves of SN 2013ej and show how the models with dense CSM improve the fit. We also compare our best-fitted model with those reported in earlier work \citep{MorozovaV_etal2017} to show how our model is more physically realistic and that the model better describes the observed data over a longer timescale. A preliminary version of this work has been presented at the IAU Symposium 331 \citep{Das_etal2017}.

\section{Methods and Simulations}
\textcolor{black}{Early spectroscopic observations of SN 2013ej in its host galaxy NGC 628(M74) confirmed it as a young type-II supernova \citep{2013ej0}. Photometric and spectro-polarimetric observations classified it as a type II-P supernova with unusually strong early-time polarization \citep{2013ejpol}.} Later it has been marked as a type II-L supernova because of the relatively faster decline (1.74 mag/100 days) of light curve in intermediate phase \citep{Bose_etal2015}. We simulate and study the evolution of the progenitor of SN 2013ej in three steps: the evolution since pre-main-sequence(MS) to post-MS till $\sim$20 years before collapse, the last two decades till few ($\sim$ 2-3) years before collapse and last 2-3 years. The final state is exploded to investigate the optical and NIR light curves.\\
\begin{figure}
\includegraphics[trim=4 3 8 5,clip, width=0.475\textwidth, height=5.25 cm]{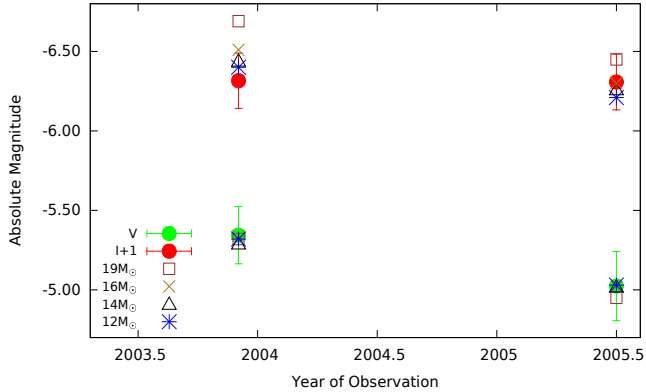}
\caption{Constraint on ZAMS mass of SN 2013ej progenitor using archival images of HST(WFC). Apparent magnitude has been corrected for distance and interstellar extinction using values quoted in \citet{Bose_etal2015} to convert it into absolute magnitude. The error bars represent composite of error in photometric data and errors in estimation of distance and interstellar extinction. The circumstellar dust extinction for 12-14M$_\odot$, the mass range whose color falls within the observed limit, turns out to be $\sim$0.7-1.2 mag in V band and $\sim$0.6-0.8 mag in I band which clearly implies reddening and circumstellar extinction 3-6 times the average galactic extinction of $\sim$ 0.2 mag. Even after dust opacity correction, the I magnitude of 16 M$_\odot$ and 19 M$_\odot$ calculated by MESA fall outside the 1$\sigma$ error bars of the observed magnitude. } 
\label{fig0}
\end{figure}
\subsection{Progenitor Evolution in the late stage and Circumstellar Medium}
We have taken the metallicity (= 0.295Z$_\odot$) of the HII region near the site of explosion (region no. 197 reported in \citet{Cedres}) and ZAMS mass range of 12-19M$_\odot$ in step of 1M$_\odot$ for all of our MESA models of nonrotating single stars\footnote{\citet{Fraser_etal2014} identified a nearby source in the archival images of its progenitor. The distance between the two ($\sim$2 pc) eliminated the possibility of its being a part of a binary system.}. 
\textcolor{black}{Although the upper mass limit of type II-P SNe progenitors observed from the pre-explosion images is found to be ZAMS mass of 17M$_\odot$ \citep{mass_upper_lmt}, higher masses of red supergiants ($\sim$ ZAMS mass of 25 M$_\odot$) have been observed in Milky Way and other galaxies. Also, models of single stars between 8 and 25 M$_\odot$ leading to successful II-P/L SNe have been predicted. This has led to the so called 'red supergiant problem'.\footnote{ \textcolor{black}{However, a recent study of the initial masses of the RSG progenitors to type II-P SNe by \cite{DB}, investigating the systematic error due to large bolometric corrections (BC) using RSGs in star clusters, and accounting for finite sample sizes of direct observations of the SN progenitors raises the upper limit of progenitors to upwards of 25 M$_\odot$ (95\% confidence limit of $<$33M$_\odot$), thus eliminating any strong evidence of "missing" SN progenitors with ZAMS mass$>$ 17M$_\odot$} } So far, the hydrodynamical models of type II-P SNe \citep{upper_lmt_model1,upper_lmt_model2} have not been able to achieve a reasonably close upper limit of mass of 17M$_\odot$.}
We use Ledoux criterion for convection, and take mixing parameters following \citet{MorozovaV_etal2016}. We keep the average long term mass loss rate around $\sim$10$^{-6}$M$_\odot$yr$^{-1}$ using Vink's scheme for hot (T$_s >$ 10$^4$K) wind with $\eta$=1 and Dutch scheme for cool wind with $\eta$=1 and 0.5 \textcolor{black}{\citep{Dutch1,Dutch3,Dutch2}}. It was constrained by X-ray studies of \citet{Chak_etal2016} which probed the mass loss rate of the progenitor of SN2013ej in the range of 40-400 years before the explosion. \\
To constrain the ZAMS mass we convert the HST magnitudes (F435W, F555W and F814W) into Johnson's photometric system (B, V and I) using the algorithm proposed in \citet{Sirianni_etal2005} and compare the archival HST(WFC) images of November 2003 and June 2005 with our simulated models in MESA\footnote{\citet{Fraser_etal2014} identified a nearby source dominating in bluer wavelengths which explained the unusual B-V color of the candidate supernova. So B band magnitude, partially contaminated by the unrelated source, is not used to constrain the mass limit}.
\textcolor{black}{While the studies by \cite{Kochanek_2017} from the same observations of \cite{Fraser_etal2014} over five years (2003-2008) do not find any variability on an average, we find a somewhat noticeable reduction in V magnitude \textcolor{black}{($\Delta$m$_V$=0.32, with $\sigma\sim$0.2 mag)} and hence increase in V-I color \textcolor{black}{($\Delta$(V-I)=0.31, with $\sigma\sim$0.28 mag)} in a smaller timescale (from November,2003 to June,2005), which suggests possible temporal changes in between. Our assumption is supported by the finding of \cite{DB} who shows that RSGs may evolve in spectral types as they approach the final stage.} We relate this to the absorption and reddening by dust formed in the CSM nearby. An optimized combination of enhanced mass loss rate, duration and instant of enhanced mass loss provided the required CSM extinction (see Appendix \ref{Appendix 2}) and this put an upper limit of ZAMS mass 14M$_\odot$ (Fig.\ref{fig0}). It is consistent with the upper limit inferred from the same archival image observation of \cite{Fraser_etal2014} and nebular phase observation of \cite{Yuan_etal2016} as well. The existence of circumstellar dust, which has smaller absorption coefficient in larger wavelengths ($Q_{abs}(\lambda=0.1\mu m)/Q_{abs}(\lambda=100\mu m) = 7300$; \cite{Draine}) also substantiates the progenitor's mid-IR brightness around 2004 reported in \cite{2013ej6}.   \\
Once the ZAMS mass is constrained, the models are allowed to evolve with average mass loss rate in 'Dutch scheme' \textcolor{black}{\citep{Dutch1,Dutch3}}. The enhanced mass loss rate is triggered by hydrogen stripping and flashes with an upper limit of $\sim$1 M$_\odot$yr$^{-1}$ in the last $\sim$2-3 years of evolution 
\textcolor{black}{(see Appendix \ref{Appendix 1} for details).} Any heavy mass loss much before this would move away too far from the star and become dilute to substantially affect the early light curves of the explosion. We use networks of 21 isotopes during heavy element fusion, and keep Fe core infall velocity limit of 10$^6$m/s as stopping condition of pre-SN evolution.  
\begin{figure}
\subfigure[] 
{
\includegraphics[trim=2 0 12 5,clip, width =0.5\textwidth, height= 5.25 cm]{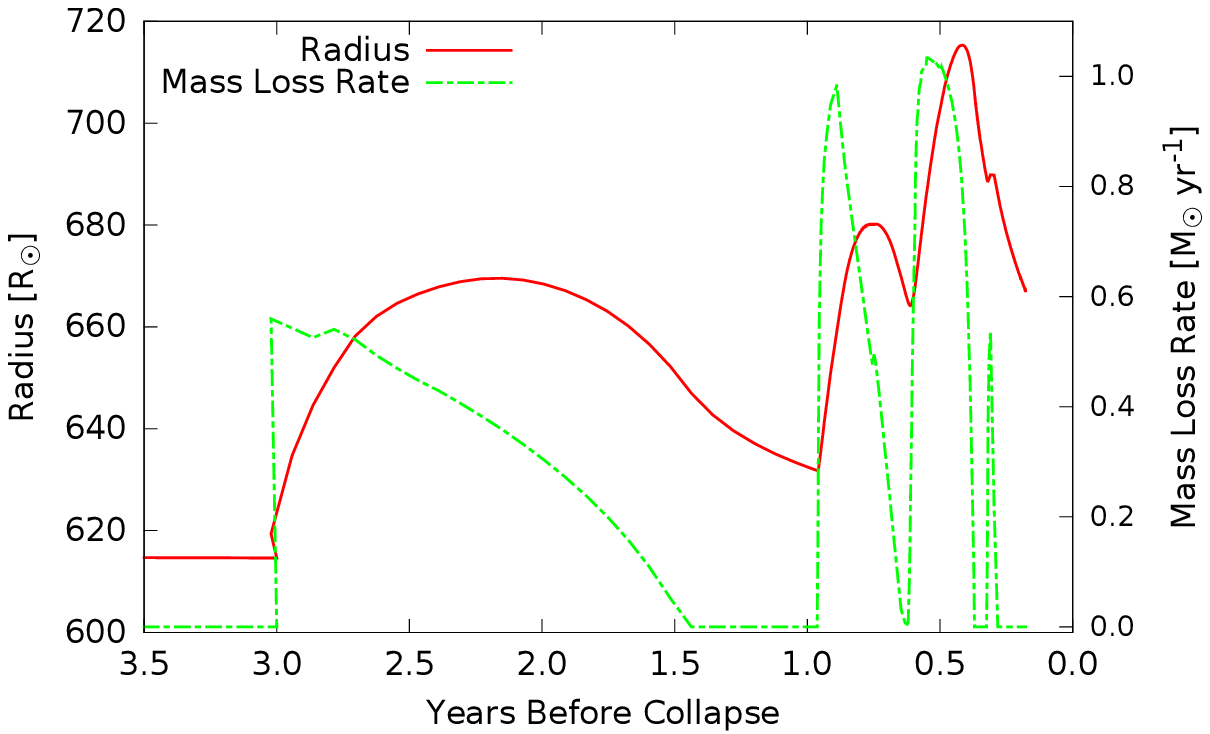}
}
\subfigure[]
{
\includegraphics[trim=2 0 12 4,clip, width =0.5\textwidth, height= 5.25 cm]{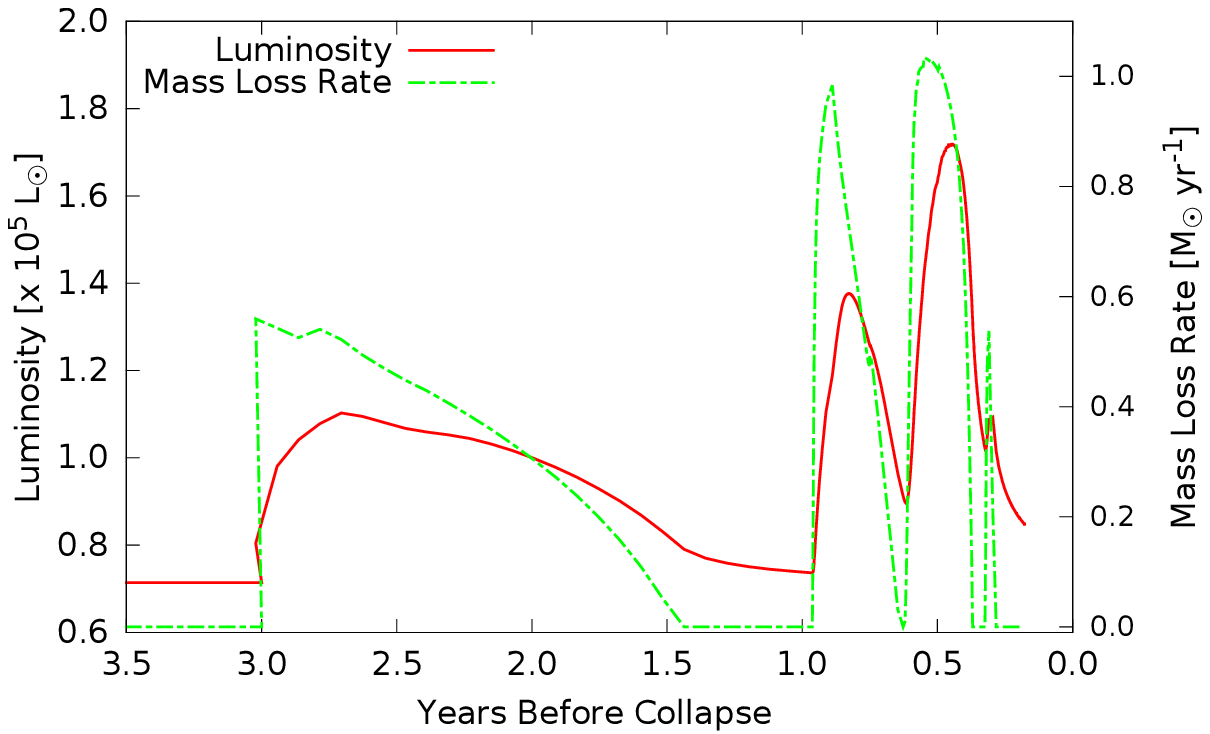}
}
\caption{Variation in radius (top) and luminosity (bottom) with large mass loss rate in late stage of stellar evolution of a model of ZAMS mass 13 M$_\odot$.} 
\label{fig1}
\end{figure}
\\Noting that the bolometric luminosity and radius vary with mass loss rate (Fig. \ref{fig1}), we calculate the CSM profile from the history of mass loss rate, radius and surface temperature assuming asymptotic wind speed of 10 km/s. Although the launch velocity was found to vary with mass loss rate, we assume the ejected gas has had enough time to relax to the asymptotic speed during the episodic ejection. We do not apply any mechanical, hydrodynamic or thermodynamic interaction between adjacent shells of ejected gas. The only source of thermal evolution is assumed to be adiabatic cooling due to expansion. The chemical composition and electron fraction in CSM are taken to be the same as that of the surface. In contrast to a smooth continuous inverse-square radial profile of density, our constructed CSM from the episodic mass loss (see Appendix \ref{Appendix 3}) is a collection of discrete shells of non-monotonically varying density (see Fig. \ref{fig2}). 
\subsection{Explosion Characteristics} \label{sec:SNEC}
We have stitched the resultant CSM to the progenitor and exploded via SNEC using a thermal bomb\footnote{Since thermal bomb and piston produce similar outputs but the former takes less computational time as pointed out by \cite{MorozovaV_etal2015}, we have used thermal bomb all through}. 
\begin{figure}
 \subfigure[] {
\includegraphics[trim=0 0 10 8,clip, width =0.45\textwidth, height= 5.25 cm]{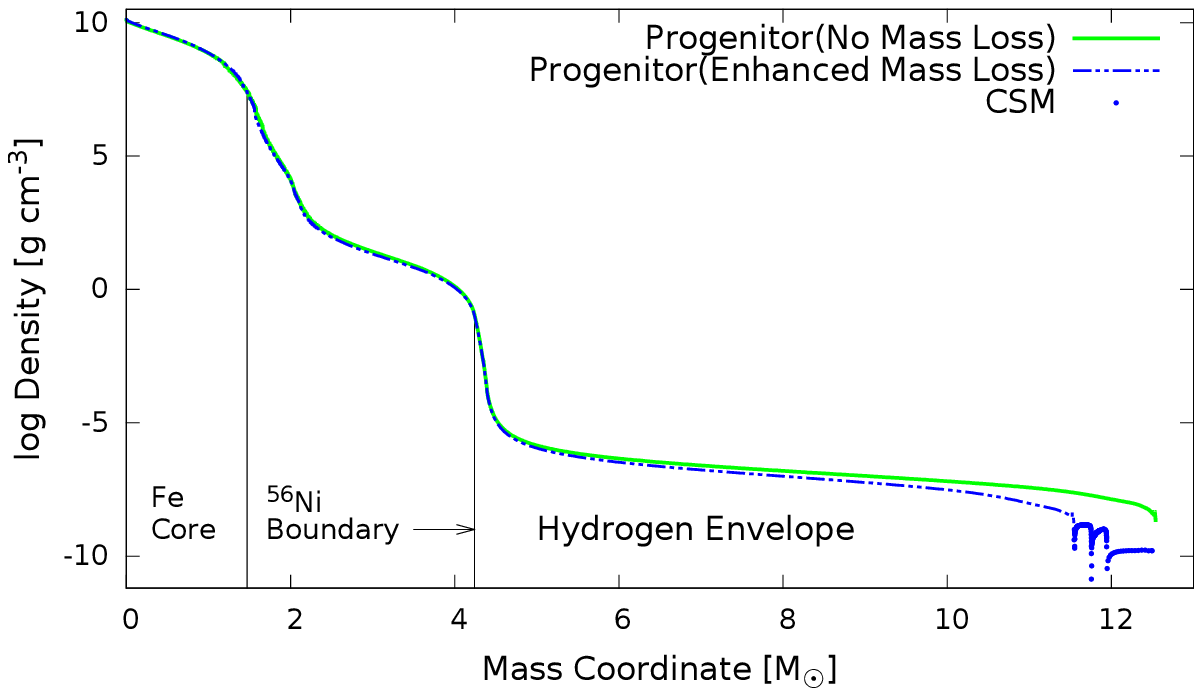}
}
\subfigure[] 
{
\includegraphics[trim=0 0 10 6,clip, width =0.45\textwidth, height= 5.25 cm]{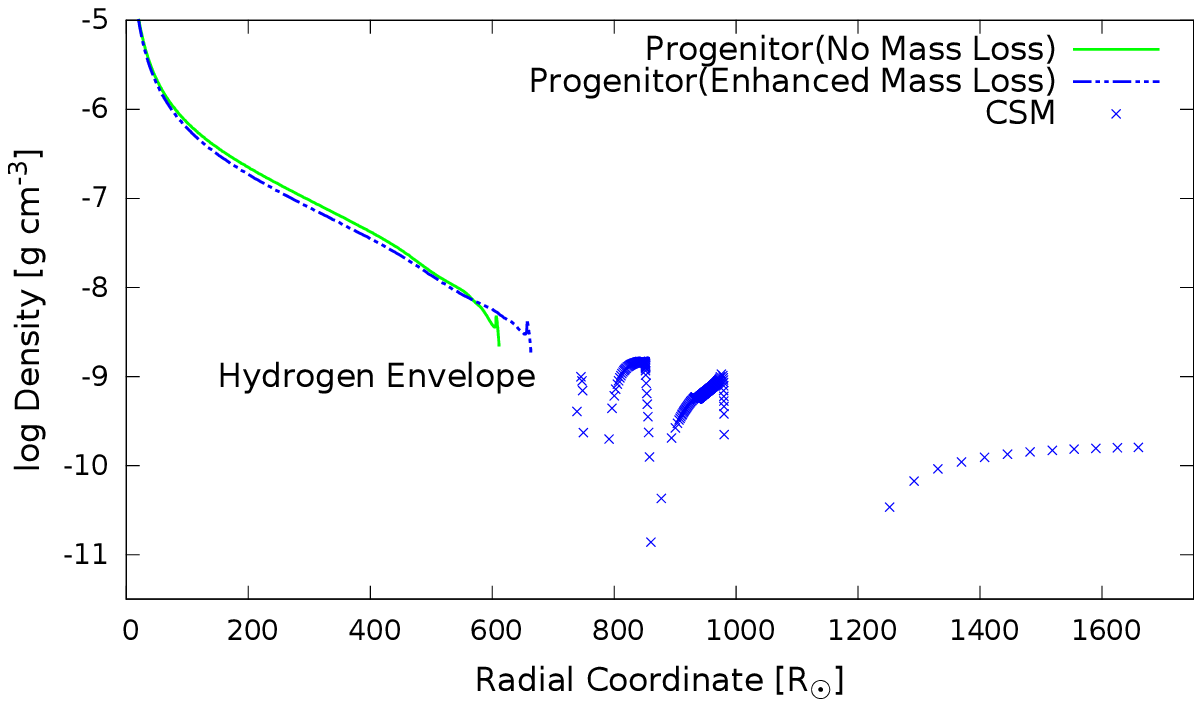}
}
\caption{Pre-SN density profile of a progenitor of ZAMS mass 13 M$_\odot$. The model without enhanced mass loss and the one with shells of CSM have been shown in same plot for comparison. The radial profile (bottom) of the central part of $\sim 4.5$M$_\odot$ has been excluded to show the CSM with better resolution.} 
\label{fig2}
\end{figure}
We have varied the explosion energy in the range of $ 0.4-1.4 \times 10^{51}$ ergs in step of $0.2\times 10^{51}$ ergs. 
\textcolor{black}{We fix $^{56}$Ni mass at 0.0207M$_\odot$ \citep{Valenti_etal2016} and evenly spread it from outside the Fe core to the middle of He and C core of each progenitor.}
After boxcar smoothing this distribution gets modified with a tail extending upto the He core boundary (Fig. \ref{fig2}a). Since $^{56}$Ni is injected externally in the simulation, it does not have any velocity distribution appropriate for the explosion. Initial spatial distribution of $^{56}$Ni over mass coordinate as referred to above is designed in a way such that it can mimic the shocked distribution as found in 3-D simulations involving explosive nucleosynthesis (Right panel, Fig. 6, \cite{Wong_etal2017}). The mass of the remnant has been taken to be equal to Fe core mass of each progenitor, which is typically $\sim$1.4-1.5M$_\odot$. Since density profile has a steep fall and velocity profile has a turnover from homologous compression ($v\propto -r$) to gravitational infall ($v \propto -{\frac{1}{\surd r}}$) across the boundary of Fe core, and shock energy is a sensitive function of kinetic energy and gravitational potential energy, we choose the remnant mass carefully. This is in contrast with earlier works where $^{56}$Ni is spread upto a constant mass coordinate and the mass of the remnant is also a constant irrespective of the internal structure of the progenitor. The light curves have been calculated till 120 days to cover a part of the radioactive tail after plateau. The magnitude in different bands have been calculated using proper bolometric corrections with blackbody approximation.\\
\section{Results}
The resultant light curves in g, V, R, I and z bands\footnote{Since SNEC does not calculate Fe line blanketing in U and B bands, these bands are excluded from goodness of fit analysis. SN2013ej was observed also in the r and i bands \citep{Valenti_etal2014}. Since r and i bands overlap with the wavelength range of V, R and I bands, considering only gVRIz bands serves our purpose of covering a range from g to z.} have been compared with the data given by \cite{Rich2014} and \cite{Yuan_etal2016}. The reduced error of our SNEC outputs against the data has been defined following \cite{MorozovaV_etal2017} as 
\begin{equation}
\chi_{\nu}^2 ={\sum_{\lambda \in[g,V,R,I,z]}\sum_{t_i \leq t_{pl}} { (m_{obs}(t_i,\lambda) - m_{th}(t_i,\lambda))^2 \over \sigma_i^2}} \\
             \times {1\over(\sum_{\lambda \in [g,..,z]} {n_\lambda} -5)}
\label{eq: chisquare}
\end{equation}
Here $t_{pl}$ is plateau length which is $\sim$ 99 days\footnote{By fitting the V-band light curve with the function $y(t) = {{{-a_0} / {(1+{exp{{t-t_{pl}}\over w_0}})}} + (p_0\times t) + m_0}$ as proposed in \cite{Valenti_etal2016}, the best fit value of the plateau duration is $t_{pl}$= 98.77 days. Here t is the time from explosion in days, a$_0$ is the depth of the drop i.e. the transition between the plateau and radioactive phases, while w$_0$ measures the slope of the drop. The parameter p$_0$ constrains the slope before and after the drop.}, $m_{th}$ = $M_{th}$ + (distance and extinction correction) where $M_{th}$ is the absolute magnitude returned by SNEC\footnote{In our analysis we have used different values of distance and extinction correction obtained in previous studies. So, to test both of the predicted absolute
magnitude and the correction factors against the observed values, we keep
observational data as a fixed point of reference and use the absolute magnitude from MESA
to calculate the "theoretical" value of apparent magnitude $m_{th}$.},  $n_\lambda$ is no. of data points in a particular band. All sources of errors i.e. the photometric error in observational data, uncertainty in distance and extinction estimation have been included in $\sigma$\footnote{$\sigma_\lambda^2 = \sigma_{m_{obs}(\lambda)}^2 + \sigma_{dM}^2 + \sigma_{A_\lambda}^2 + abs(M_{th}(\lambda))$. Here 'dM' stands for distance modulus. Assuming a Poisson distribution of simulated values, we take $\sigma_{{th}(\lambda)}^2 = abs(M_{th}(\lambda))$}. We have used the distance and interstellar extinction correction quoted in \textcolor{black}{\cite{Rich2014}, \cite{Bose_etal2015}, \cite{Huang_etal2015} and \cite{Yuan_etal2016}}. We find that for the set of values quoted in \textcolor{black}{\cite{Rich2014}} (distance d=9.12$\pm$0.84 Mpc and A$_V$= 0.21$\pm$0.04) the best fitted model with CSM has ZAMS mass 12- 13M$_\odot$ and explosion energy $E_{exp}$ = $0.6\times 10^{51}$ergs, while the values of \textcolor{black}{\cite{Yuan_etal2016}} and \textcolor{black}{\cite{Bose_etal2015}} (distance d=9.57$\pm$0.7 Mpc and A$_V$= 0.185$\pm$0.004) predict best fitted ZAMS mass = 13M$_\odot$ and a range of $E_{exp}$ = $0.6-0.8 \times 10^{51}$ergs. \textcolor{black}{\cite{Huang_etal2015}} has considered a larger value of extinction by taking the reddening of host galaxy M74 into account. Using their values (distance d=9.6$\pm$0.5 Mpc and A$_V$= 0.37$\pm$0.19) we can pinpoint a ZAMS mass 13M$_\odot$ and $E_{exp}$ = $0.8\times 10^{51}$ergs. 
\begin{figure}
\subfigure[]
{
\includegraphics[trim=10 12 0 33,clip, width =0.5\textwidth, height= 5.25 cm]{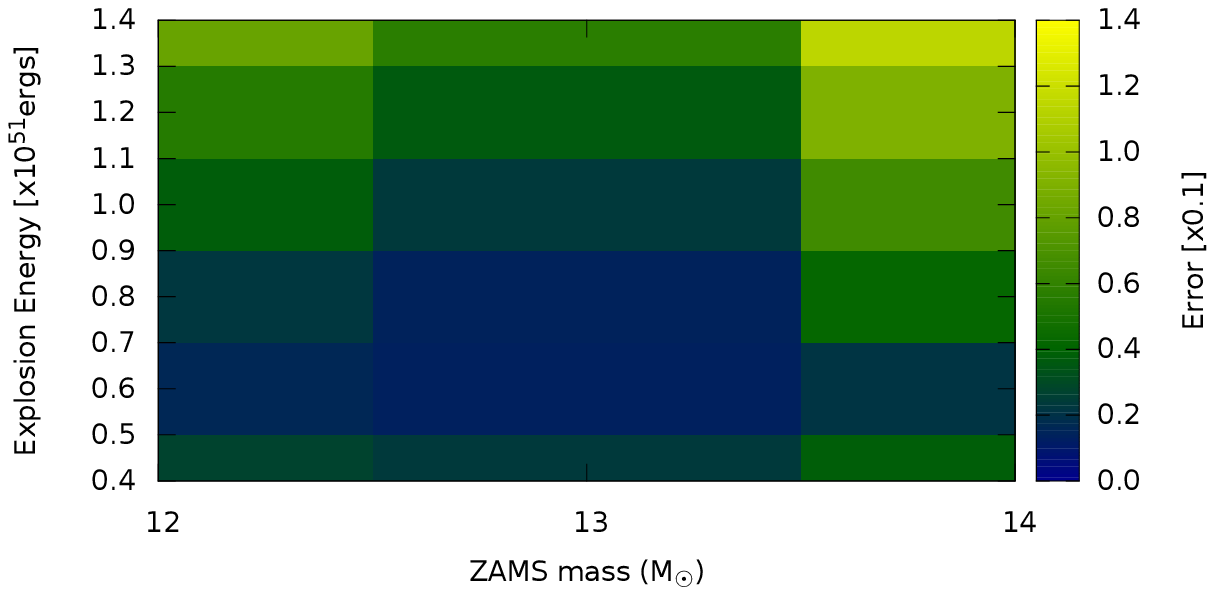}
}
\subfigure[]
{
\includegraphics[trim=10 14 0 33,clip, width =0.5\textwidth, height= 5.25 cm]{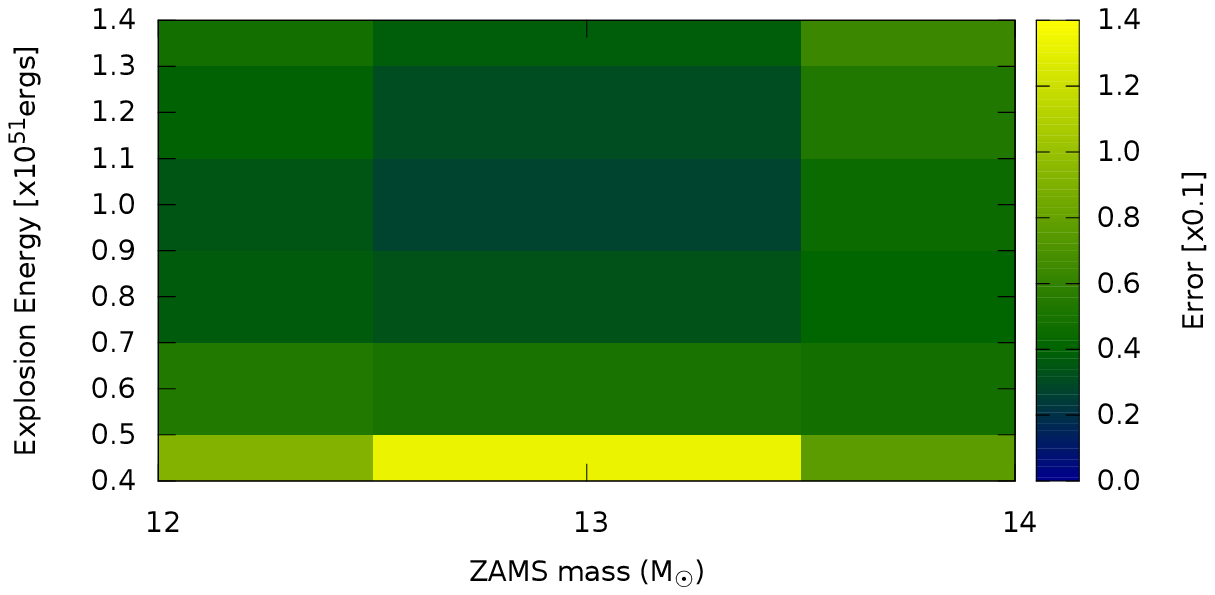}
}
\caption{Error maps of the models with CSM(top) vs without CSM(bottom) of different ZAMS mass and explosion energy against g, V, R, I and z bands of SN 2013ej (data from \citet{Yuan_etal2016}). The photometric error in observational data, errors in estimation of the distance of host galaxy and average galactic extinction (taken from \citet{Huang_etal2015}) have been included in the error analysis.} 
\label{fig3}
\end{figure}
The light curves of models without dense nearby CSM deviate from the data with a different distribution (Fig. \ref{fig3}); they have a goodness of fit worsened typically by a factor of ~2 compared to the light curves of explosions taking place in dense, nearby circumstellar medium (Table \ref{tab1}). We find that the required explosion energy for the progenitors with dense CSM is reduced by almost a factor of 2 compared to the case without any dense and compact CSM. \textcolor{black}{The results have been cross-checked against the data from \cite{Rich2014, Valenti_etal2014} and \cite{Bose_etal2015} to confirm that the best fitted ZAMS mass and explosion energy turn out to be similar from analysis with all datasets.} We see that the model with CSM best fitted using Huang's values of the distance and interstellar extinction correction is visually closest to the data (Fig. \ref{fig4}). It strengthens the consideration of host extinction which is confirmed by recent analysis of massive star population around the site of explosion \citep{Maund2017}. Although it was intended to fit upto the plateau only, since SNEC is not reliable beyond that region \citep{MorozovaV_etal15}, the model with dense and nearby CSM fits the radioactive tail reasonably well in four (gVRI) bands. Also, the model predicts the peak magnitude in U and B bands correctly. On the other hand, the best-fit model without CSM hardly satisfies any feature of the light curve.
\begin{figure}
 \includegraphics[trim=4 0 5 10,clip, width=0.5\textwidth, height= 6.0 cm]{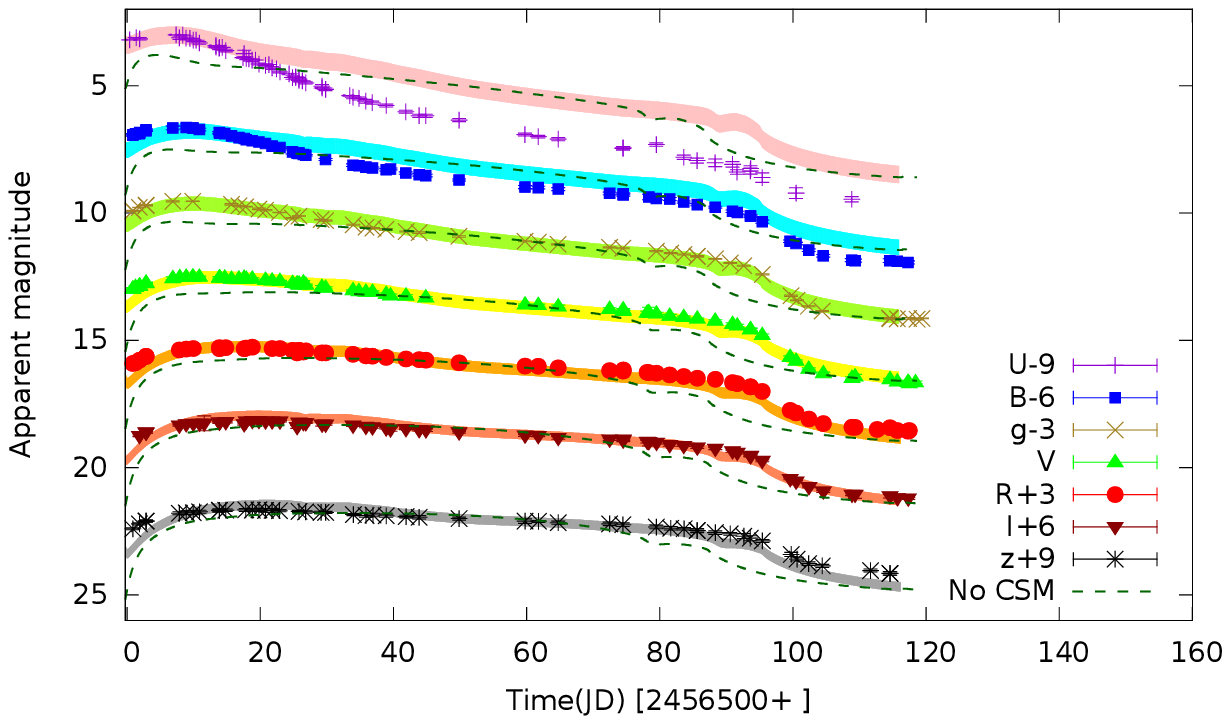} 
 \caption{The NUV(U)-optical(BgVR)-NIR(Iz) light curves of SN 2013ej (data: \citet{Yuan_etal2016}). The curves for best-fitted model (fitted against gVRIz bands till the plateau of $\sim$ 99 days) with CSM are shown with the errors at each point. Values of distance and extinction estimation have been taken from \citet{Huang_etal2015} with relative extinction relation borrowed from \citet{Cardelli}. Our best-fitted model without CSM is shown in dashed lines for the sake of comparison.}
\label{fig4}
\end{figure}
From this we can infer that the progenitor of SN 2013ej was very likely to be surrounded by dense compact slowly moving nearby CSM which was formed due to enhanced mass ejection in very late stage of its evolution. This analysis is consistent with the spectroscopic observations of \cite{Bose_etal2015} where a weak ejecta-CSM interaction was inferred from the high velocity $H\alpha-H\beta$ profiles.\\
We also compare our result with earlier works \textcolor{black}{\citep{MorozovaV_etal2017}} and find that our model (with CSM) is better reproducing the data over a longer period. Both models match the early light curves (first $\sim$20 days) equally well, but the one from \textcolor{black}{\cite{MorozovaV_etal2017}} starts overestimating the light output after $\sim$50 days. On the other hand, our model reproduces data very well in the whole plateau region, during the transition from plateau to radioactive phase and also in the beginning of radioactive phase (Fig. \ref{fig5}). 
\begin{table*}
  \begin{center}
  \caption{Overview of progenitor and explosion properties best-fitted for SN 2013ej}
  \label{tab1}
  \begin{tabular}{c c c c c c c}\hline 
{\bf ZAMS} & {\bf Pre-SN} & {\bf CSM } & {\bf Pre-SN } & {\bf CSM } & {\bf Energy} & {\bf Goodness of Fit$^5$} \\
{\bf mass [M$_\odot$]} & {\bf mass [M$_\odot$]} & {\bf mass [M$_\odot$]$^1$} & {\bf radius [$R_\odot$]} & {\bf radius [$R_\odot$]$^2$} & {\bf [$10^{51} $ergs]$^3$} & {[\bf$\chi_{\nu}^2$]}  \\ \hline
 13 & \textcolor{black}{11.602} & \textcolor{black}{0.964} & 667 & 1650 & 0.6-0.8 & 0.0013-0.0014 \\ 
 13 & \textcolor{black}{12.566} & ---  & 617 & --- & 1.0-1.2 & 0.0027-0.0031 \\ \hline
  \end{tabular}
 \end{center}
\vspace{2mm}
\begin{flushleft}
 \footnotesize{
 {\it Notes:}\\
  $^1$ lost mass during enhanced mass loss rate in last few ($\sim 2-3$) years\\
  $^2$ the external radius of the dense CSM formed in the last few years before explosion \\
  $^3$ asymptotic energy of the shock after breakout\\
  $^4$ $^{56}$Ni mass was kept fixed at 0.0207M$_\odot$ \textcolor{black}{\citep{Valenti_etal2016}}\\
  $^5$ The range of explosion energy corresponds to the range of $\chi_{\nu}^2$, which is treated as a region of valley instead of a single point of minima because of almost same $\chi_{\nu}^2$ value over the range of energy. Note the better goodness of fit for the first row of the Table}
\end{flushleft}
\end{table*}
\begin{figure}
 \includegraphics[trim=4 0 5 10,clip, width=0.5\textwidth, height= 6.0 cm]{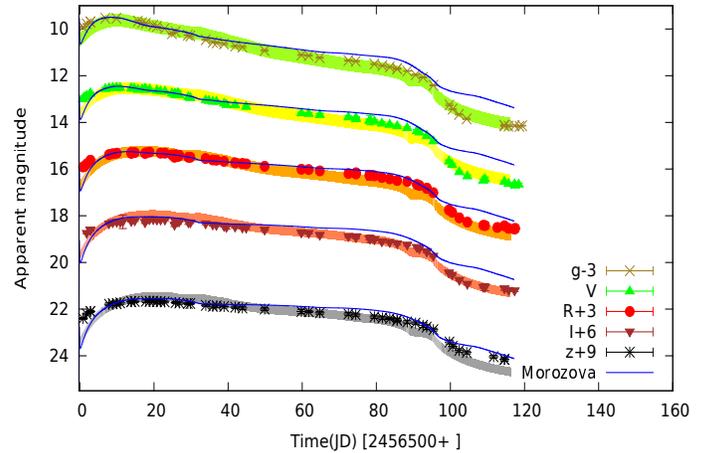} 
\caption{Comparing our work with \textcolor{black}{\citet{MorozovaV_etal2017}} against optical(gVR) and NIR(Iz) light curves of SN 2013ej (data: \citet{Yuan_etal2016}). \textcolor{black}{Note how our work differs from their result after $\sim$50 days and better represents the data.}   }
   \label{fig5}
\end{figure}
\section{Summary and Discussions}

In this paper we have discussed how the same ZAMS star but with different mass loss scenarios can pass through two evolutionary tracks differing in a timescale much less than the Kelvin-Helmholtz timescale and end up as different pre-SN progenitors, although occupying roughly the same position in H-R diagram. The density, velocity and temperature profiles of hydrogen envelope of the stars with no huge mass loss history are substantially different from those of the progenitors surrounded by dense and compact CSM (Fig.\ref{fig2}); even though total mass (envelope + CSM) is same in both cases. So the rise and plateau of the light curve, which are functions of hydrogen profile are expected to be different. In these cases, although the pre-SN images of the progenitor star may not be distinguishable unless the star is very nearby, the light curves of resulting explosions will differ substantially. We note some particular differences due to the presence of dense and compact CSM: (1) UV and optical bands are simultaneously bright in early phase, and (2) for a fixed shock energy the optical light curve is brighter at peak, flatter after peak and the plateau is longer. These are consistent with what was noticed and discussed by \cite{Moriya_etal2011} in a simpler scenario with artificially designed smooth and continuous wind approximation. The simultaneous brightness in NUV and optical can be explained by the recombination and de-excitation of elements in the CSM which are (respectively) ionized and excited by the decelerated shock passing through the CSM. A larger fraction of the shock energy transferred to the dense CSM gets thermalized and eventually diffusively radiated. It brightens the early light curve and flattens the light curve after peak by reducing the adiabatic loss. From the plateau length it is clear that the same hydrogen mass spread over a larger radius contribute more towards the light output. The key difference between this scenario and usual type II-n SNe is that the extended CSM in type II-n is optically thin, while here the photosphere lies in the compact CSM for upto $\sim$40 days since explosion. In usual type II-n the interactive phase lasts for $\sim$1 yr but in our case the intense interaction happens for $\sim$ 2-3 days and the breakout takes place when the shock has already traversed through the dense and compact CSM. The CSM behaves like a gravitationally unbound extended part of the progenitor, but is definitely different from an inflated stellar envelope. Since the lack of detection of early radio emission and flash-ionization in the CSM points to the proximity and compactness by putting constraint on the radial extent of the CSM \citep{Yaron_etal2017}, our assumption of dense and compact CSM is supported by the null radio emission from SN 2013ej in an early VLBI observation by \cite{2013ej5}. \\
We also account for extinction carefully in pre-SN stage and show how the dust opacity can affect the pre-SN images. Unless calculated correctly, this may lead to wrong inferences of the properties of the progenitor which can get propagated into further analysis. The CSM which contributed in supernova light curves did not need any similar opacity corrections. Before the explosion, the dense circumstellar material could not move far away from the star to cool down to form dust steadily. Also, when the shock comes out of the star and hits the CSM it deposits a larger fraction of its energy to the CSM, part of which heats the CSM up. This will sublimate all dust even if it is formed earlier.\\
Along with confirming earlier studies \citep{MorozovaV_etal2017, Moriya_etal2017} that nearby wind-like circumstellar material indeed affects early light curves of SNe, we show that the late-time enhanced mass ejection can result in such dense and nearby CSM. \textcolor{black}{Our method and results support the suggestion and explanation by \cite{Quat2012, SmithN_etal2014,Moriya2014} and \cite{Fuller_2017} that the mass loss rates of red supergiants  increase by several orders of magnitude prior to the SN explosion. There are different mechanisms proposed to explain late stage mass loss enhancement based on the increase in nuclear power in the core. The basic principle is generation of a vigorous convective motion leading to internal gravity waves carrying energy from the core to hydrogen envelope, which either damp through shock and diffusive radiation and steepens pressure near the stellar surface or convert into sound wave which dissipates near the stellar surface resulting in outbursts or heavy mass loss. The final progenitors along with CSM may be capable of explaining the diversity in type II-P/L SNe or intermediate of II-P and II-n SNe.}    
We argue that the better performance of our model indicates that the light curves depend on the profile of CSM and the progenitor, and hence our approach of letting the star pass through the phase of increased mass loss rate and constructing CSM from that stellar history is physically reasonable. We note the in-phase change in radius and bolometric luminosity with heavy mass loss rate (e.g. Fig. \ref{fig1}). This indicates that pulsation may have connections with late outburst and eruptions which are not usually included in standard stellar evolution codes. Ideally, the simulation should allow the star to naturally lose mass at any rate depending on a particular phase of post-MS evolution. The combination of the impact of mass loss variation on the star and the CSM formed as a consequence leads to a detailed profiling of the explosion input. Sampling a range of parameters related to mass loss and advanced modeling of the compact and nearby CSM might help in explaining the qualitative similarity of types II-P and II-L SNe and the finer details of spectral output of each category and these questions are under investigation.                

\section*{Acknowledgements}
Research reported here was a part of the 12$^{th}$ Five-Year Plan project on "Astrophysics of Supernovae and Neutron Stars" (12P-0261) at Tata Institute of Fundamental Research (TIFR). S.D. thanks the Visiting Students' Research Program (VSRP) of TIFR for the opportunity to work on this project. A.R. acknowledges the Raja Ramanna Fellowship at Homi Bhabha Centre for Science Education (TIFR). We thank the anonymous referee for a careful reading of the manuscript and for comments that helped in improving our presentation.

\textcolor{black}{\software{MESA-r8118 \citep{Pax2011,Pax2013,Pax2015}, SNEC v1.01 \citep{MorozovaV_etal15}, GNUPLOT v4.6}}

\bibliographystyle{yahapj}

\appendix

\section{MESA input models}\label{Appendix 1}

A sample inlist of the post MS evolution is shown below. If not explicitly mentioned, all other parameters have been kept at their default values.

\texttt{\\ 
star$\_$job \\ 
! start a run from a saved model \\
load$\_$saved$\_$model = .true.\\
saved$\_$model$\_$name = '13M$\_$z0.006.mod'\\ 
history$\_$columns$\_$file = 'history$\_$columns$\_$postMS.list'  \\
profile$\_$columns$\_$file = 'profile$\_$columns$\_$postMS.list' \\
!nuclear reaction network \\
adv$\_$net = 'approx21.net'\\ 
!end of star$\_$job namelist\\ 
\& controls \\ 
! use C/O enhanced opacities  \\ 
use$\_$Type2$\_$opacities = .true.  \\
Zbase=0.006 \\     
! mixing  \\
mixing$\_$length$\_$alpha = 2\\
use$\_$Ledoux$\_$criterion = .true. \\
alpha$\_$semiconvection = 0.1 \\
overshoot$\_$f$\_$above$\_$nonburn$\_$core = 0.025 \\
overshoot$\_$f0$\_$above$\_$nonburn$\_$core = 0.05\\
overshoot$\_$f$\_$above$\_$nonburn$\_$shell = 0.025\\
overshoot$\_$f0$\_$above$\_$nonburn$\_$shell = 0.05\\
overshoot$\_$f$\_$below$\_$nonburn$\_$shell = 0.025\\
overshoot$\_$f0$\_$below$\_$nonburn$\_$shell = 0.05\\
overshoot$\_$f$\_$above$\_$burn$\_$h$\_$core = 0.025\\
overshoot$\_$f0$\_$above$\_$burn$\_$h$\_$core = 0.05\\
overshoot$\_$f$\_$above$\_$burn$\_$h$\_$shell = 0.025\\
overshoot$\_$f0$\_$above$\_$burn$\_$h$\_$shell = 0.05\\
overshoot$\_$f$\_$below$\_$burn$\_$h$\_$shell = 0.025\\
overshoot$\_$f0$\_$below$\_$burn$\_$h$\_$shell = 0.05\\}
\\
$\alpha_{mlt}$ is mixing length parameter, $f_{ov}$ and $f_0$ are exponentially decreasing overshooting (both for the core and the convective shells), and $\alpha_{sc}$ is semi-convection efficiency. We do not change the default option of thermohaline mixing. \\ \\
\texttt{! configure mass loss of RGB \& AGB\\
cool$\_$wind$\_$RGB$\_$scheme = 'Dutch' \\
hot$\_$wind$\_$scheme = 'Vink'\\
cool$\_$wind$\_$AGB$\_$scheme = 'Dutch' \\
RGB$\_$to$\_$AGB$\_$wind$\_$switch = 1d-4\\
Vink$\_$scaling$\_$factor =1d0\\
Dutch$\_$scaling$\_$factor = 1d0	\\
super$\_$eddington$\_$scaling$\_$factor = 1\\
!artificial halt \\
! max$\_$age=17659996.33 \\ }
\\
Since episodic mass loss mechanisms are not incorporated in the code, the stellar evolution is artificially halted at different points by using different value of maximum age of our choice less than the lifetime of the star and thus mass loss rate can be tuned in selected intervals. \textcolor{black}{Instead of running a grid of the \textit{enhanced mass loss rate} and its \textit{duration}, we have considered possible constraints on the \textit{duration of the enhanced mass loss} (duration$\sim$ the difference of age at final core-collapse and the initiation time of the mass loss enhancement)). The former two are in some ways coupled and have together substantial effects on the SN light curves when the progenitor star explodes. Since luminosity of the star goes up with mass loss increase, and \cite{Fraser_etal2014} has progenitor observation till 2008 without any trace of remarkable increase in brightness, the enhanced mass loss could not have taken place more than 5 years before the final collapse. Moreover, the null result from early radio observations by \cite{2013ej5} suggest that the dense CSM, if it exists, should be compact and close to the progenitor. Now, most of the predictions of late stage mass loss enhancement is based on the increase in nuclear energy production rate \citep{Quat2012, Moriya2014,Fuller_2017}. So we try to relate the \textit{duration of the enhanced mass loss} with the time when two major reactions : $^{16}$O and $^{28}$Si burning stages dominate the energy production. 
\begin{figure}[h]
\centering
\includegraphics[scale=1.25]{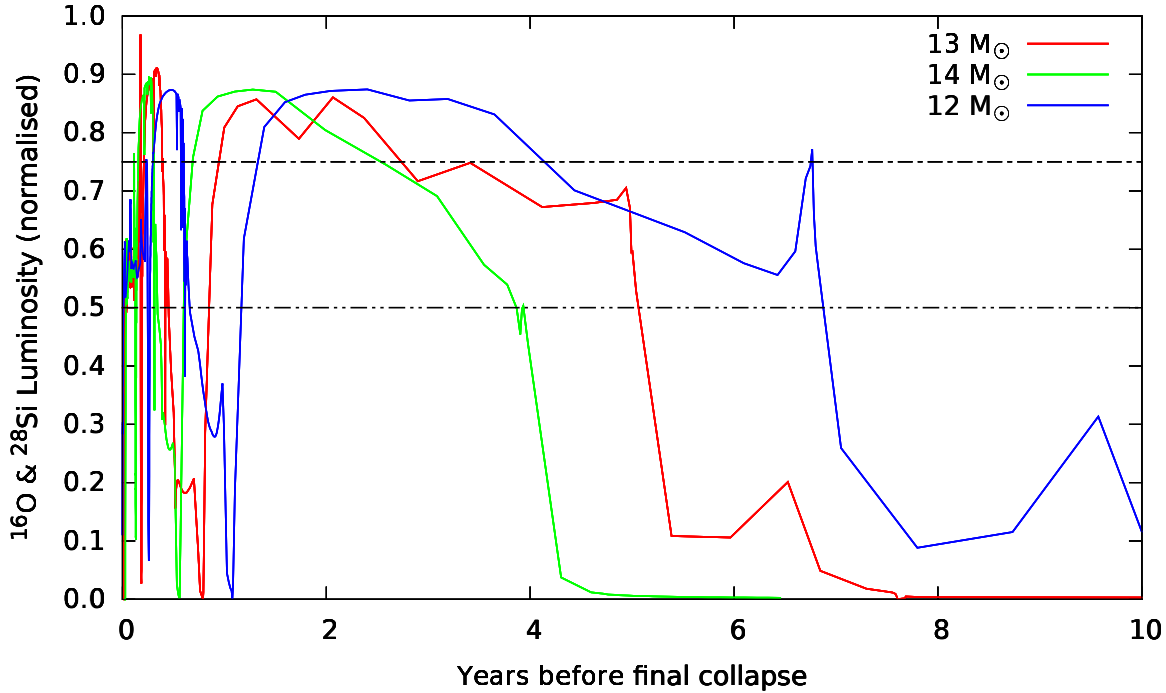}
\caption{Increase of O+Si Burning Power (normalized wrt total burning power which includes core and all shell burning zones) in late stage of post-MS stellar evolution}
\label{fig6}
\end{figure}
In Fig. \ref{fig6} we see that for 12-14 M$_\odot$ (ZAMS) stars the energy production rate becomes more than 75$\%$ of total nuclear luminosity $\sim$2.5-4.5 years before the final collapse, which suggests the mass loss should get enhanced around this timescale. We tried with mass loss enhancement 1 year before the collapse, but the actual mass loss rate never became as high as 1 M$_\odot$yr$^{-1}$ even though the upper limit was increased to $\sim$3-10 M$_\odot$yr$^{-1}$. This is probably because this timescale is not enough for the star to jump to a high mass loss rate ($\sim$1 M$_\odot$yr$^{-1}$) and relax back to equilibrium. Then we chose mass loss enhancement 2-3 years before the collapse, which worked satisfactorily. So, choosing other values of the \textit{duration of the enhanced mass loss} after taking into account of all constraints stated above would only change the final progenitor profile by a factor of few, if at all.}\\ \\
\texttt{ !enhanced H loss in last 2 years \\
remove$\_$H$\_$wind$\_$mdot = 1d0 \\
flash$\_$wind$\_$mdot =  0.375d0 \\ }
\\
\textcolor{black}{Apart from the usual schemes of standard mass loss, hydrogen stripping and flashes are two ways to artificially trigger any change in mass loss. In case of hydrogen stripping, hydrogen is ejected
(i.e. 'stripped') from the envelope with large momentum. In MESA one can only put an upper limit on the rate of mass loss, and the hydrogen stripping goes on until some stopping conditions
are met. On the other hand, flashes can emit the mass in a shorter timescale which is
usually in the same order of the temporal resolution of the particular model,
and can reach the limit of mass loss mentioned by the user. Since the physical explanation of the enhancement in late stage mass loss of massive stars are debatable, the proposed mechanisms in the literature are not incorporated in MESA. The commands mimic the effect of outburst/flare observed in late stages of stellar evolution.}
\\
Smaller values (0.01 M$_\odot$yr$^{-1}$) have been tried $\sim$ 15-20 years before the final collapse to study the opacity effect of dust in the CSM. To study the effect of dense and compact CSM on the early light curves of SNe, larger values of mass loss rate (1.0 M$_\odot$yr$^{-1}$) were triggered. \textcolor{black}{We consider 1 M$_\odot$yr$^{-1}$ as as order of magnitude estimate of high mass loss rate. Since this is an upper limit and not the exact value, few multiples or fraction of it would not necessarily lead to an noticeably different mass loss history and consequently different CSM profile, and therefore would not change the SN light curves substantially when the shock interacts with the dense circumstellar medium.} Since this enhancement continued for 2-3 years only, we did not put any lower limit of hydrogen mass as a stopping condition.\\ \\
\texttt{
!resolution\\
mesh$\_$delta$\_$coeff$\_$for$\_$highT = 1.0\\ }
\\
Near the center of the star mass resolution is increased three times the default value.\\\\
\texttt{ log$\_$directory = '13M$\_$z0.006$\_$w1'\\
! tuning  interval \\ 
profile$\_$interval = 5 \\
history$\_$interval = 5 \\
photostep = 2 \\}
\\
The 'photostep' is reduced to a small value to save the intermediate models at smaller time intervals so that the evolution can be restarted with new set of parameters from any instant of our choice.\\
\\
\texttt{min$\_$timestep$\_$limit = 1d-9 \\
max$\_$timestep = 0.05d7} 
\\
\\The temporal resolution is increased by reducing maximum timestep during the episodes of enhanced mass loss rate.

\section{Calculation of dust extinction in circumstellar medium}\label{Appendix 2}

\begin{figure}
\centering
\includegraphics[scale=1]{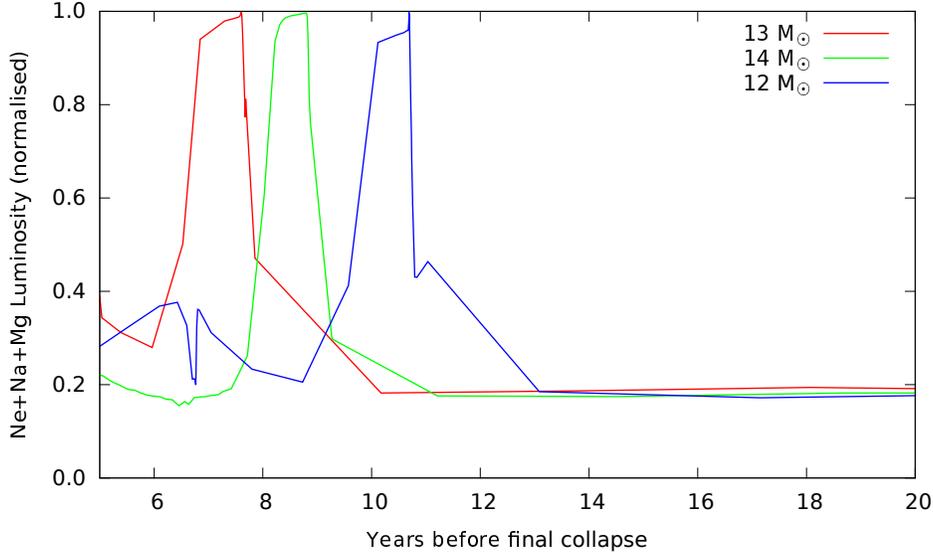}
\caption{Increase of Ne+Na+Mg Burning Power (normalized wrt total burning power which includes core and all shell burning zones) in late stage of post-MS stellar evolution}
\label{fig7}
\end{figure}

\begin{figure}
\centering
\includegraphics[scale=1]{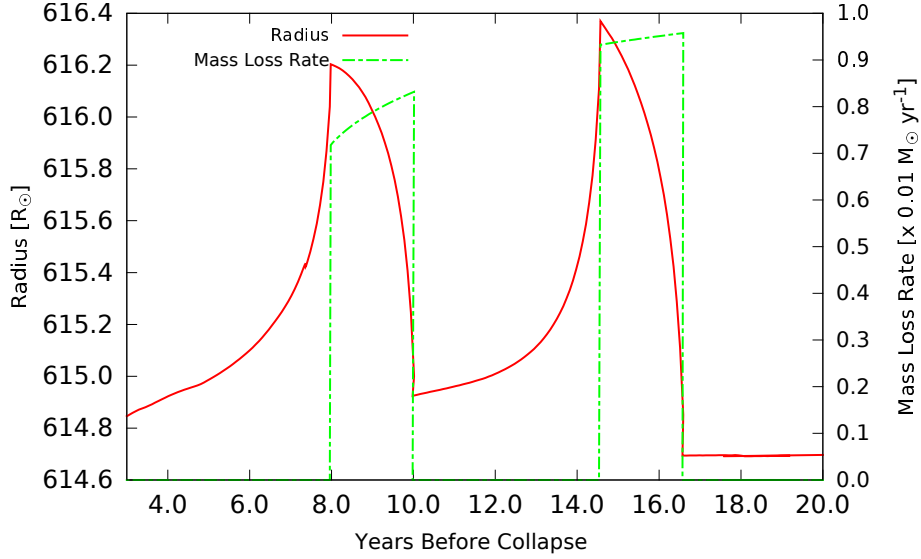}
\caption{Pulsation during mass loss enhancement of ZAMS 13 M$_\odot$ star}
\label{fig8}
\end{figure}

After the mass is ejected from the star it moves away, cools down and forms dust out of the elements (commonly Carbon and Silicon) available. The dust in a dense CSM can considerably affect the observed luminosity and color of the star by absorption and reddening. Following \cite{Kochanek} we borrow the interstellar dust extinction calculation from \cite{Perna} and extend it to CSM with few necessary corrections. At a given frequency, the dust extinction in a region is  
\begin{equation} 
A(\nu)= \int\limits^{R_{max}}_{R_{min}} \,dr \int\limits^{a_{max}}_{a_{min}} \,da \pi a^2{dn\over{da}}(Q_{abs}(a,\nu)+(1-g)Q_{sct}(a,\nu))
\label{eq: dust1}  
\end{equation} 
where, size distribution of dust   
\begin{equation}
{dn\over{da}} = 0.00045A n_H a^{-\beta}    ;a_{min} \leq a \leq a_{max}
\label{eq: dust2} 
\end{equation}
We took following approximations and corrections: 
\begin{enumerate}
\item From the composition of the surface of the star, we find that only graphite dust can be formed in the CSM. We assume that $50\%$ of the total $^{12}C$ has formed dust. Density of dust grains $\rho_{dust}$ has been been kept fixed at the average value of 2.26 gm/cc.
\item Since MESA is a 1-D calculation, angle dependent $Q_{sct}(a,\nu)$ of equation \ref{eq: dust1} has been neglected. 
\item Since CSM is close to the star and hence illuminated by the stellar radiation we take $\beta$= 1.81 (eqn. \ref{eq: dust2}) from \cite{Perna}. This is different from the galactic value of 3.5 under an undisturbed medium approximation. 
\item Since under the influence of radiation larger dust grains fragment into smaller ones and a portion of smaller grains evaporate, the range of grain radius \textit{'a'} (eqn. \ref{eq: dust1}) evolves. We have considered the asymptotic distribution of $a_{max}=$0.22$\mu$m and $a_{min}=$0.15$\mu$m with normalization constant A modified to 0.00045$\times$A (see eqn. \ref{eq: dust2}) signifying reduction in the total no. of dust particles.    
\item In the above mentioned range of grain radius we see from Fig. 4a of \cite{Draine} that coefficient of absorption $Q_{abs}(a,\nu)$ is independent of \textit{'a'}. Considering the frequency range of Johnson's B, V and I bands we find $Q_B$ = 1.4, $Q_V$ = 1.5 and $Q_I$ = 1.0.  
\item we convert no. density ($n_H$) into mass density ($\rho_H$) and express the radial integration of equation \ref{eq: dust1} in terms of distance covered by the wind. Thus the integral becomes direct function of the parameters related to mass loss: mass loss rate, wind speed, duration of mass loss , etc., which has been read and used from MESA outputs. 
\item Since the dust destruction temperature is $T_d$= 1500 K, $R_{min}$ is chosen at a point where temperature falls below $T_d$. We have assumed simple adiabatic cooling of the wind due to expansion.   
\end{enumerate}We find that the extinction at a particular time of observation not only depends upon a) the total mass lost but also on b) the \textit{duration of mass loss} and c) \textit{instant of mass loss}. The lost mass should have enough time before observation to move away from the star and cool below the dust destruction temperature, but should not move so far away that it becomes too dilute to contribute in the extinction. Secondly, the duration of mass loss will control the compactness and radial profile of CSM density on which extinction is strongly dependent. An optimized combination of mass loss rate and these two parameters can give the required extinction. 
\textcolor{black}{We tuned these three variables in MESA and found that for two episodes of mass loss, roughly 1.5-2.0 years duration, and with an upper limit of $\sim$0.01 M$_\odot$yr$^{-1}$, one of which occurred $\sim$15 years before final collapse (i.e. $\sim$5 years before the 2003 HST observation) and another that occurred $\sim$10 years before collapse (i.e. between 2003 and 2005), a CSM is formed with a mass of $\sim0.05$ M$_\odot$ spread over $\sim$1000R$_\odot$. This can give rise to an extinction of 0.7-1.2 mag in V band. We attribute the mass loss enhancements to small pulsation (${{\Delta R} \over R} < 1\%$) and the increase of Ne (and Na,Mg) burning power (see Fig \ref{fig7} and \ref{fig8}) as found from MESA outputs.}

\section{Construction of circumstellar profile}\label{Appendix 3}

We have constructed the CSM similar to conventional RSG winds. Density of an ejected shell of gas is defined as   
\begin{equation}
\rho_{wind} = {\dot{M} \over {4\pi r^2 v_{wind}}}
\end{equation}In our work $r= v_{wind}\times t + R_{star}(t)$, '\textit{t}' being the time before final collapse, and R$_{star}$ is the radius of the star when the gas was ejected. We assume that the gas is emitted from the surface of the star. So in a given MESA output, where the variables were the \textit{upper limit of mass loss rate} and \textit{the duration of the enhanced mass loss}, we do not have any free parameter as such. The external radius of the CSM \textit{R$_{ext}$} is primarily determined by the maximum value of '\textit{t}' such that $\rho_{wind}$ does not fall below $\sim 10^{-11}$ g cm$^{-3}$, i.e. $\sim$1$\%$ of the stellar surface density. \textcolor{black}{Maximum value of '\textit{t}' is in the order of \textit{the duration of the enhanced mass loss}. Since $\dot{M}$ is a time dependent variable in our simulation, and R$_{star}$ was found to vary with $\dot{M}$ (see Fig.\ref{fig1}), the density profile depends strongly on the history of mass loss and surface properties of the star, and does not follow any constant integer-power radial dependence as opposed to the assumption of \cite{Moriya_etal2012} and \cite{Moriya_etal2017} (see Fig.\ref{fig2})}. Although the surface temperature (L = 4$\pi$R$_{star}^2\sigma$T$_s^4$) hardly changes during the period of episodic ejection, the temperature profile of the CSM does not remain smooth or constant because of its radial dependence. Since temporal resolution was increased during enhanced mass loss, the enclosed mass in CSM is calculated as $M= \sum \dot{M}(t_i)\times \Delta t_i$ using simple trapezoidal rule. We had $\sim$260 data points to sum over $\sim$2-3 years. 

\section{Explosion parameters in SNEC} \label{Appendix 4}
The final profile of our MESA models have been fed as inputs of SNEC. Since the convention of indexing cells in SNEC is opposite from MESA (from surface to center), the outputs of MESA have been flipped and arranged in a format compatible with SNEC before use. Here we actually skip the intermediate detail of the core bounce and formation of shock. The calculation in SNEC starts from the moment when the shock has gained enough energy to come out of the infalling matter of the star and started its final outward propagation through the rest of the star. The main explosion parameters used here in SNEC are explosion energy, $^{56}$Ni mass and its spread over mass coordinate, and the mass of the remnant which would eventually turn into a neutron star or black hole. Explosion energy is the asymptotic energy of the shock once it comes out of the stellar ejecta and CSM in the vicinity, if any. Since SNEC always assumes a successful explosion, the initial shock energy is calculated as the sum of binding energy (kinetic+ internal+ gravitational energy of the stellar material the shock will pass through) and user given explosion energy. Radioactive $^{56}$Ni is artificially injected to mimic the effect of explosive nucleosynthesis. $^{56}$Ni mass reflects the slope of the exponential tail after plateau. Its spread over the stellar mass determines how fast it would get exposed to the photosphere and directly contribute in the optical light curves. SNEC does not account for any post-shock fallback mass, so the mass coordinate from where the shock starts moving outward is equal to the mass of the remnant. For a given stellar model we have varied explosion energy only. The detail of how other parameters are chosen is mentioned in section \ref{sec:SNEC}.             
\end{document}